\documentclass[preprint2]{aastex631}

\usepackage{amssymb,amsmath,graphicx,natbib,hyperref}
\usepackage{booktabs}
\usepackage{longtable}
\usepackage{multirow}
\usepackage{url}
\usepackage{graphicx}
\usepackage{color}

\newcommand{\Ha}{$\rm{H}\alpha$}

\DeclareRobustCommand{\ION}[2]{%
\relax\ifmmode
\ifx\testbx\f@series
{\mathbf{#1\,\mathsc{#2}}}\else
{\mathrm{#1\,\mathsc{#2}}}\fi
\else\textup{#1\,{\mdseries\textsc{#2}}}%
\fi}
\newcommand{\kms}{km\,s$^{-1}$}

\newcommand{\nii}{[\ION{N}{ii}]}
\newcommand{\oi}{[\ION{O}{i}]}

\newcommand{\oiii}{[\ION{O}{iii}]}
\newcommand{\sii}{[\ION{S}{ii}]}

\newcommand{\ha}{H$\alpha$} 
\newcommand{\hb}{H$\beta$}

\newcommand{\XS}{$\mathtt{XS}$}

\shorttitle{Non-axisymmetry in NGC~1087}
\shortauthors{Lopez-Coba et al.}

\begin{document}

\title{Unveiling a hidden bar-like structure in NGC~1087: kinematic and photometric evidence using MUSE/VLT, ALMA and JWST } 

\correspondingauthor{C. Lopez-Coba}
\email{calopez@asiaa.sinica.edu.tw}

\author[0000-0003-1045-0702]{Carlos~L\'opez-Cob\'a}
\affiliation{Institute of Astronomy and Astrophysics, Academia Sinica, No. 1, Section 4, Roosevelt Road, Taipei 10617, Taiwan}
\author[0000-0003-1045-0702]{Lihwai~Lin}
\affiliation{Institute of Astronomy and Astrophysics, Academia Sinica, No. 1, Section 4, Roosevelt Road, Taipei 10617, Taiwan}
\author[0000-0001-6444-9307]{Sebasti\'an~F.~S\'anchez}
\affiliation{Instituto de Astronom\'ia, Universidad Nacional Autonoma de M\'exico, Circuito Exterior, Ciudad Universitaria, Ciudad de M\'exico 04510, Mexico}

\begin{abstract}
We report a faint non-axisymmetric structure in NGC\,1087 through the use of JWST Near Infrared Camera { (NIRCam)}, with an associated kinematic counterpart observed as an oval distortion in the stellar velocity map, \ha~and CO~$J=2\rightarrow1$ velocity fields. This structure is not evident in the MUSE optical continuum images but only revealed in the near-IR with the F200W and F300M band filters at $2\mu$m and $3\mu$m respectively.
Due to its elongation, this structure resembles a stellar bar although with remarkable differences with respect to conventional stellar bars. Most of the near-IR emission is concentrated within $6\arcsec~\sim500$~pc with a maximum extension up to 1.2~kpc. The spatial extension of the large-scale non-circular motions is coincident with the bar, which undoubtedly   
confirms the presence of a non-axisymmetric perturbation in the potential of NGC\,1087. The oval distortion is enhanced in CO due to its dynamically cold nature rather than in \ha. We found that the  kinematics in all phases including stellar, ionized and molecular, can be described simultaneously by a model containing a bisymmetric perturbation; however, we find that an inflow model of gas along the bar major axis is also likely. Furthermore the molecular mass inflow rate associated can explain the observed star formation rate in the bar.
 This reinforces the idea that bars are mechanisms for transporting gas and triggering star formation. This work contributes to our understanding of non-axisymmetry in galaxies using the most sophisticated data so far.
\end{abstract}

%%%%%%%%%%%%%%%%%%%%%%%%%%%%%%%%%%%%%%%%%%%%%%%%%%

%%%%%%%%%%%%%%%%% BODY OF PAPER %%%%%%%%%%%%%%%%%%

\section{Introduction}

 Stellar bars or ``bars'' are one of the most visual signs of non-axisymmetry in galaxies \citep{3RC}. Like other morphological structures in galaxies, bars are not well defined, although they are relatively easy to identity in composite images because of their bar-shape structure \citep[e.g.,][]{Masters2011,Cheung2013}, and their prominent bar dust lanes \citep{Athanassoula1992}.    
 NGC\,1087 is a clear example where the combination of high spatial resolution data allowed us to reveal a hidden bar that is only observed in the near-infrared (NIR).
 The James Webb Space Telescope (JWST) is allowing us to reveal structures not seen before due to the lack of resolution and sensitivity. Their infrared bands are prone to detect the emission from old-stellar structures like bars $\sim10$~Gyr \citep[][]{sanchez-blazquez11}.
 In the optical, dust obscuration can prevent their detection, affecting the global statistics of galaxies hosting bars in the nearby Universe \citep[e.g.,][]{sell93}.
 { Apart} from their optical and NIR characteristics, bars leave imprints of non-axisymmetry in the kinematics of gas and stars \citep[e.g.,][]{Wong2004, fathi05, LopezCoba2022}; thus studying the dynamical effects of bars at local scales is crucial for understanding the role they play in galaxy evolution.
 A comprehensive study of bars could be addressed if kinematics and wide-range photometric observations were accessible for a considerably large sample of galaxies. Yet, spatial resolution plays an important role { in separating} the different structural components of galaxies.
 { For example, the large integral field spectroscopic surveys like CALIFA \citep{sanchez12a} or MaNGA \citep{ManGA}, although providing a large statistical sample of galaxies, their nominal resolution of $2\farcs5\sim1$~kpc inhibits a detailed study of individual galaxies, while biasing the sample towards bar-lengths larger than the FWHM resolution.}
 In a similar manner NIR photometric and optical catalogs provide resolution of few arcseconds which again limits the detection of small bars \citep[][]{S4G}.

 In this work, we make use of the most sophisticated data to unravel the detection of a bar in NGC\,1087. This paper is structured as follows: in Section~2 we describe the data and data analysis;  in Section~3 we address the detection of a faint stellar bar in the infrared while their ionized and molecular counterpart are addressed in Section~4; Section~5 describes the oval distortion and in Section~6 we present the discussion and conclusions.

\section{Data and Data analysis}

NGC~1087 is an intermediate Sc galaxy located at 14 Mpc \citep[e.g.,][]{Kourkchi2017}; at this distance the physical spatial resolution is 68~pc$\mathrm{~arcsec^{-1}}$. 

This work is based on public data from the Multi-Unit Spectroscopic Explorer \citep[MUSE,][]{Bacon2010}; data from the JWST Near Infrared Camera (NIRCam); and ALMA
CO~$J=2\rightarrow$1 observations, in particular we used the data products from the PHANGS--ALMA survey \citep[e.g.,][]{Leroy2021a,Leroy2021b}, namely moments 0 and 1.
The MUSE-VLT data was obtained as part of the recent release of the MUSE-PHANGS datacubes \citep[e.g.,][]{PHANGS_doi, Emsellem2022}. MUSE is an integral field spectrograph which provides spatially resolved spectra over a $1\arcmin\times1\arcmin$ field of view (FoV), covering the optical spectrum, with a spectral resolution at full width at half maximum (FWHM) of $\sim2.6$\AA. The estimated spatial resolution of the  $2\arcmin\times3\arcmin$ MUSE mosaic covering NGC~1087 is not worse than $0\farcs9$/FWHM \citep[e.g.,][]{Emsellem2022}.
We used fully calibrated data products from JWST from the JWST Science Calibration Pipeline version 1.10.1 \citep[e.g.,][]{bushouse}.
Specifically, we made use of the PHANGS-JWST first result products \citep[e.g.,][]{PHANGS-JWST_doi}, namely, the NIRCam F200W filter (FWHM $\sim0\farcs05$ and $0\farcs031/\mathrm{pixel}$ sampling), which has its nominal wavelength at $1.99\mu$m, in addition to the F360M and F335M band filters to trace the stellar continuum. Finally the spatial resolution of the ALMA data is $1\farcs60$/FWHM following \citet{Leroy2021a}.
The CO moment~0 map was transformed to surface density ($\mathrm{ \Sigma_{mol} }$), assuming a standard Milky Way $\mathrm{ \alpha_\mathrm{CO}^{1-0}=4.35\,M_{\odot} /pc^2 }$ conversion factor and $\mathrm{ \Sigma_{mol}[M_\odot pc^2] = 6.7\, I_{CO(2-1)}[K\,km/s]} \cos i $ \citep[e.g., Eq.~11 from][]{Leroy2021a}, with $i$ being the disk inclination angle estimated in $44.5^{\circ}$.

The MUSE data analysis was made with the {\sc pypipe3d} tool \citep[e.g.,][]{Lacerda2022}.
In short, {\sc pypipe3d} performs a decomposition of the observed stellar spectra into multiple simple stellar populations (SSPs) each with different age and metalicities. For concordance with \cite{Emsellem2022}, we use the same stellar libraries based on the  extended MILES \citep[E-MILES,][]{EMILES}, FWHM $=2.5$\AA;
We convolved the spectral resolution of the SSPs to that of the MUSE line-spread function \citep[e.g.,][]{Bacon2017}.
To increase the signal-to-noise (SN) of the stellar continuum, we performed a Voronoi binning-segmentation on a 2D-slice centered around the V-band, ensuring bin-sizes of the order of the MUSE-FWHM resolution and SN\,$\sim50$. This binned map will serve to compute different properties of the underlying stellar continuum, while the ionized gas properties are analyzed in a spaxel-wise sense.
The result of {\sc pypipe3d} is a set of maps comprising information about the stellar populations (stellar velocity, stellar mass, among others products), and the ionized gas (emission-line fluxes including: \ha, \hb, \sii$\lambda6717,6731$, \nii$\lambda6584$, \oiii$\lambda5007$, \oi$\lambda6300$, emission-line velocities, equivalent widths etc); see \cite{Lacerda2022} for a thorough description of the analysis and dataproducts. From this analysis we estimate the stellar mass of this object in $\mathrm{ \log M_{\star}/M_{\odot} = 9.9}$.
\begin{figure*}[t!]
  \centering
     \includegraphics[]{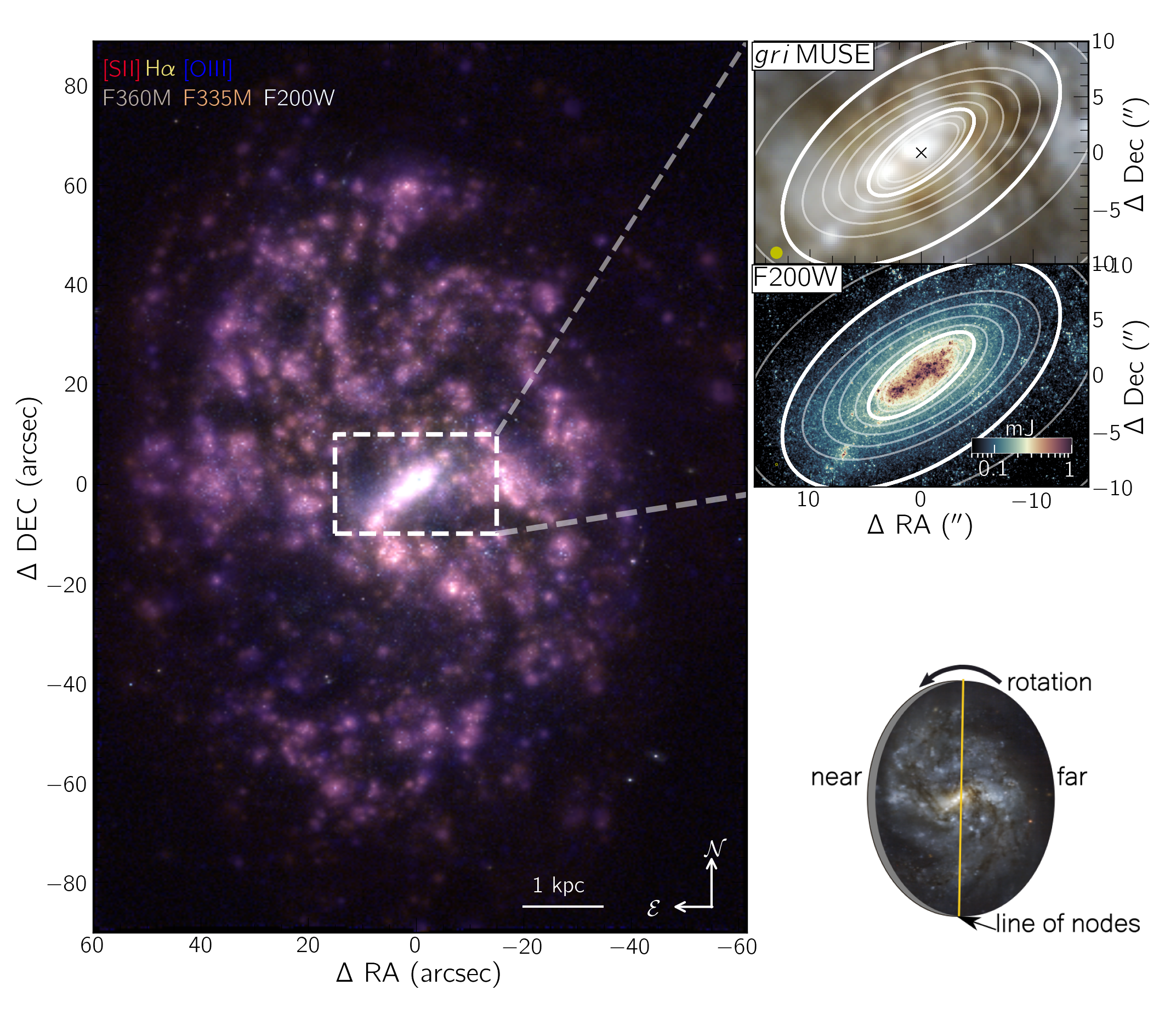}
  \caption{\textit{Central panel:} False color image showing the distribution of the ionized gas and NIR continuum in NGC~1087 with fluxes taken from MUSE-VLT and NIRCam imaging respectively (red: \sii, yellow: \ha, blue: \oiii, brown: F360M, orange: F335M, white: F200W). \textit{Right panels:} A zoom-in of the $30\arcsec\times20\arcsec$ innermost region is shown to highlight the bar-structure. The upper panel shows a true color image from the MUSE cube (R: \textit{ i}-band, G: \textit{r}, B: \textit{g}). The black cross in the middle represents the NIR nucleus. The middle panel shows the JWST-F200W image together with a set of isophotes describing the bar NIR light distribution; the isophotes show a constant orientation in the sky at $\phi^{\prime}_{bar}=304^{\circ}\pm3^{\circ}$. Two isophotes located at $6\arcsec$ and $14\arcsec$ are highlighted with thicker lines for reference. In each case the FWHM resolution is shown with a yellow circle. \textit{Bottom inset:} Orientation of the object in the sky. 
  }
  \label{fig:main} 
\end{figure*}
\section{FAINT stellar bar}

The optical continuum image of NGC~1087 exhibits a bright, fuzzy and featureless nucleus with several dust lanes as revealed by the $gri$ MUSE image in Figure~\ref{fig:main}, because of that it is unclear whether this object shows a stellar bar in the optical. Two bright spots are observed in the central region where the near-IR (NIR) nucleus is found. 
Unlike the majority of galaxies hosting stellar bars, NGC~1087 does not show a well defined bar, nucleus neither a star forming ring \citep[e.g.,][]{3RC}.
Conversely, the high spatial resolution from the F200W imaging filter allows to resolve a clear elongated and clumpy structure of presumably stellar clusters  aligned along a preferential position angle (P.A.), which differs from the disk orientation\footnote{From now on primed variables make reference to values measured on the sky plane, otherwise in the disk plane.} $\phi_{disk}^{\prime} = 359^{\circ}$, as observed in Figure~\ref{fig:main}. At $2\mu$m this structure resembles a faint stellar bar and it is obscured in the optical due to dust absorption.
F200W traces mostly stellar continuum, hence, it is expected that most of the continuum emission along this structure comes from old stars \citep[e.g.,][]{Starburst99}.
Elliptical isophotes with variable position angle and ellipticity ($\varepsilon^{\prime}$) show a preferential alignment of the $2\mu$m emission as observed in the top-right panel from Figure~\ref{fig:main}.

\begin{figure*}[t]
    \centering
    \includegraphics[]{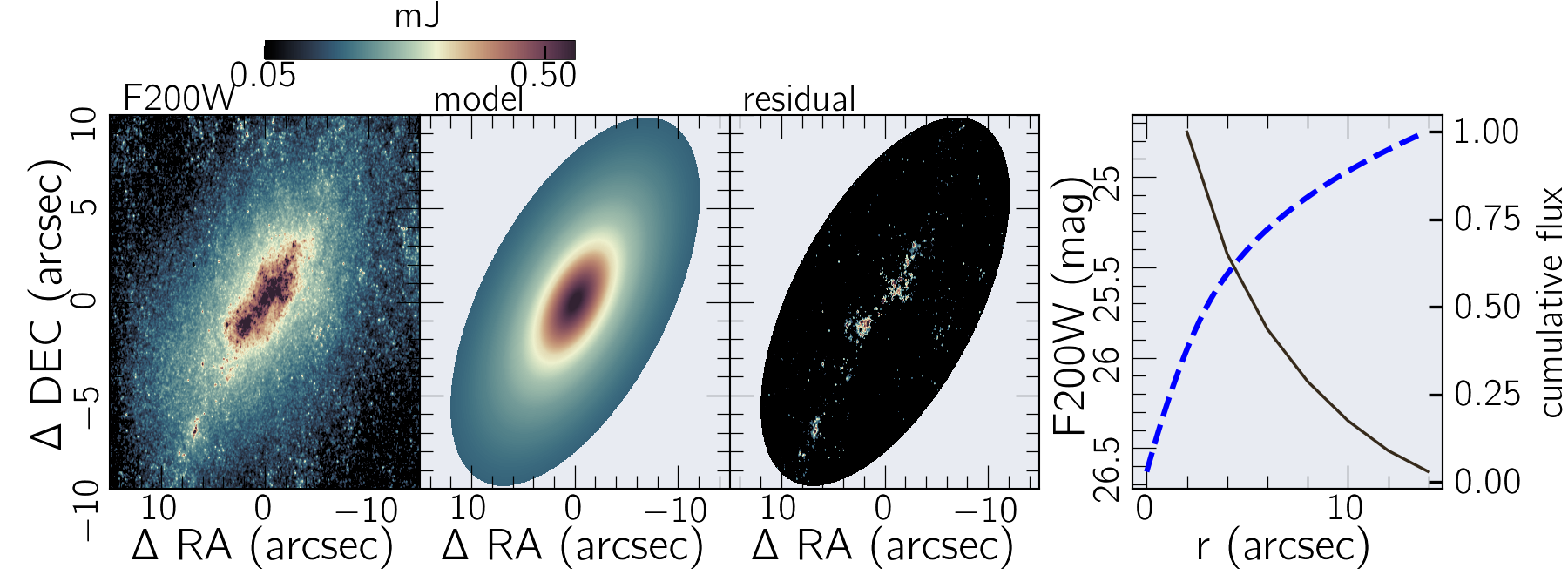}\\
    \includegraphics[]{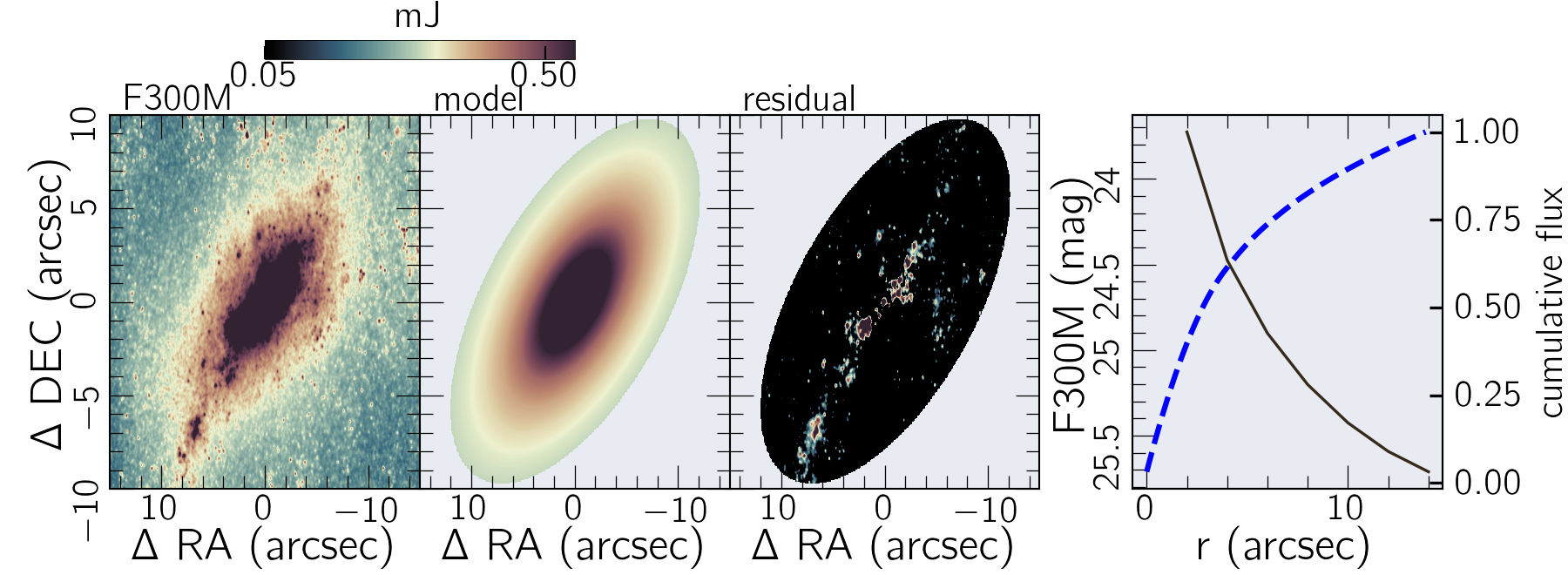}%
  \caption{NGC~1087 as observed with the F200W and F300M NIRCam band filters. {\it First panel:}  zoom around a $30\arcsec\times20\arcsec$ size window containing the bar-like structure. {\it second panel:} Best non-parametric model for the $2\mu$m light distribution using the photometric version of the \XS~code ({\it in prep.}). The coordinates of the photometric center was estimated in $\alpha_{2000} = 2^h46^m25.170^s$, $\delta_{2000} = -00^{\circ}29^{\prime}55.728\arcsec$, while the position angle of the disk describing the bar light distribution was estimated in $\phi_{bar}^{\prime} = 305^{\circ}$ with a $0.5$ ellipticity. {\it Third panel:} residuals of the modeling (observed-model). {\it Fourth panel:} the black line represents the light distribution profile of the model shown in the second panel, while the blue dashed line represents the cumulative light distribution of the model. Distances are measured following these projection angles.}
  \label{fig:xs_phot}
\end{figure*}

\begin{figure}[]
    \centering
    \includegraphics[]{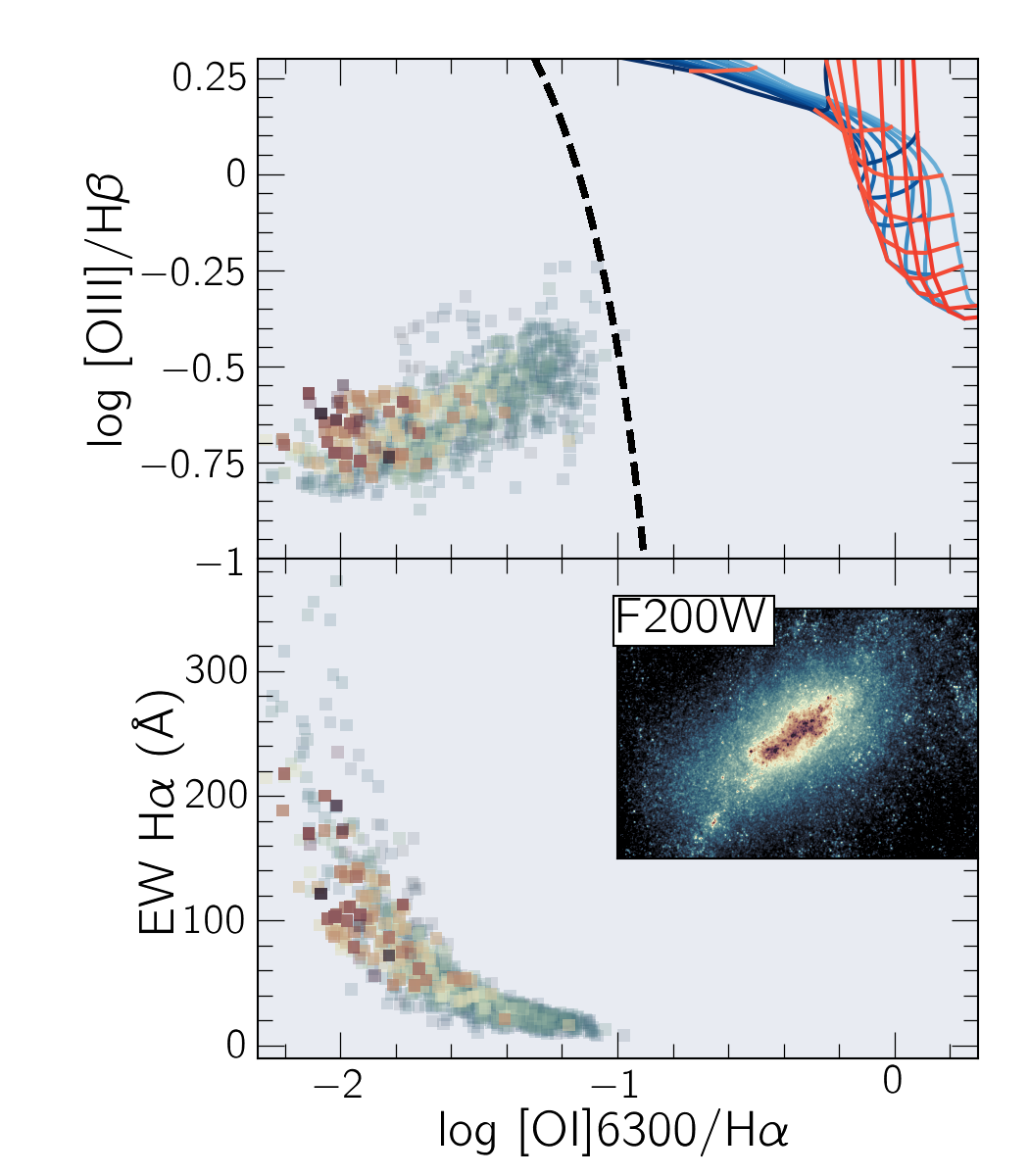}%
  \caption{Diagnostic diagrams for the ionized gas in the bar structure. {\it Top panel:} \oi6300/\ha~vs.~\oiii/\hb~diagram. The black dashed line represents the star forming demarcation curve from \cite{Kewley2006}. Shock grid models from the updated MAPPINGS-V \citep[][]{mappingsv} computed by \cite{Alarie2019} are shown for $\mathrm{n_e=10\,cm^{-3}}$, $\mathrm{Z_{ISM}=0.008~(equivalent~to ~12+ \log O/H=8.55)}$, and a wide range of magnetic fields (blue lines) and pre-shock velocities (red lines). The colors of the points map the F200W light distribution shown in the inset figure. {\it Bottom panel:} EW(\ha)~in absolute value vs. \nii/\ha~line ratio. Old and evolved stars are expected to produce EW(\ha)\,$<3$\AA.. }
  \label{fig:bpt}
\end{figure}
\begin{figure}[t!]
  \centering
     \includegraphics[]{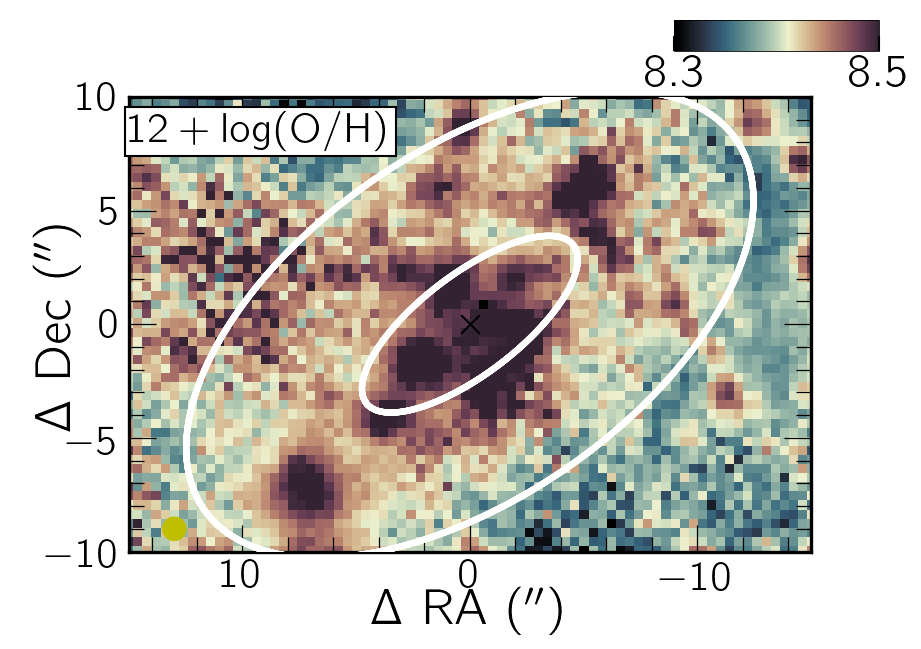}
  \caption{Oxygen abundance in the bar-region computed with the \cite{Pilyugin2016} calibrator. As in Figure~\ref{fig:main} the white ellipses describe the NIR bar. The innermost ellipse shows an average metallicity of $\mathrm{ 12+\log O/H = 8.5}$.
  }
  \label{fig:metal} 
\end{figure}

\begin{figure}[t!]
  \centering
     \includegraphics[]{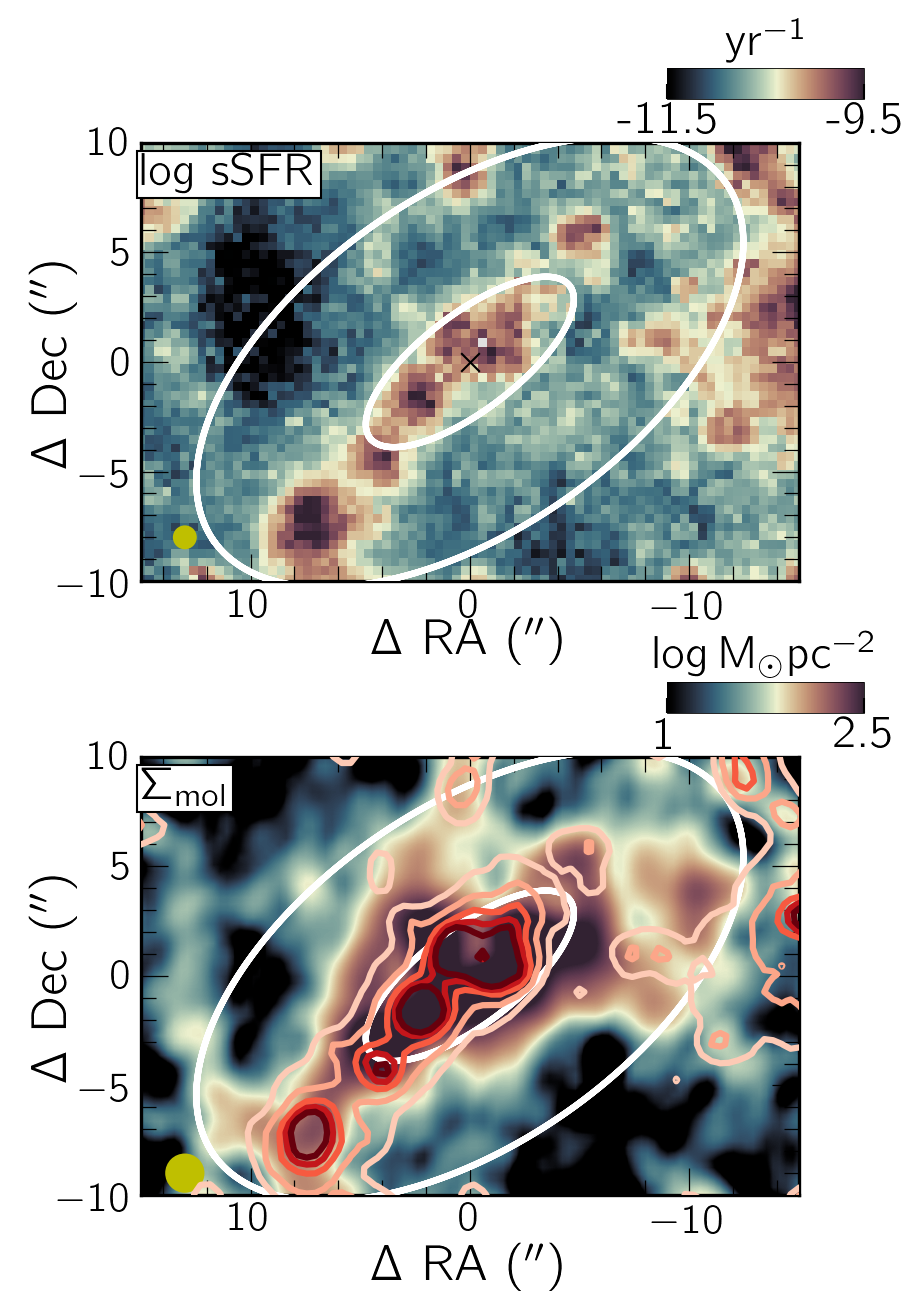}
  \caption{\textit{Top:} Specific star formation rate computed through the \ha~based SFR and the stellar mass density maps, $\mathrm{sSFR = SFR/M_{\star}}$. \textit{Bottom:} Molecular surface density around the bar structure. The \ha~flux is overlaid with reddish colors for comparison. The yellow circles represent the FWHM spatial resolution in each case. The white ellipses delineate the NIR bar extension.
  }
  \label{fig:ssfr} 
\end{figure}

\begin{figure*}
  \centering
    \includegraphics[width=\textwidth,height=!]{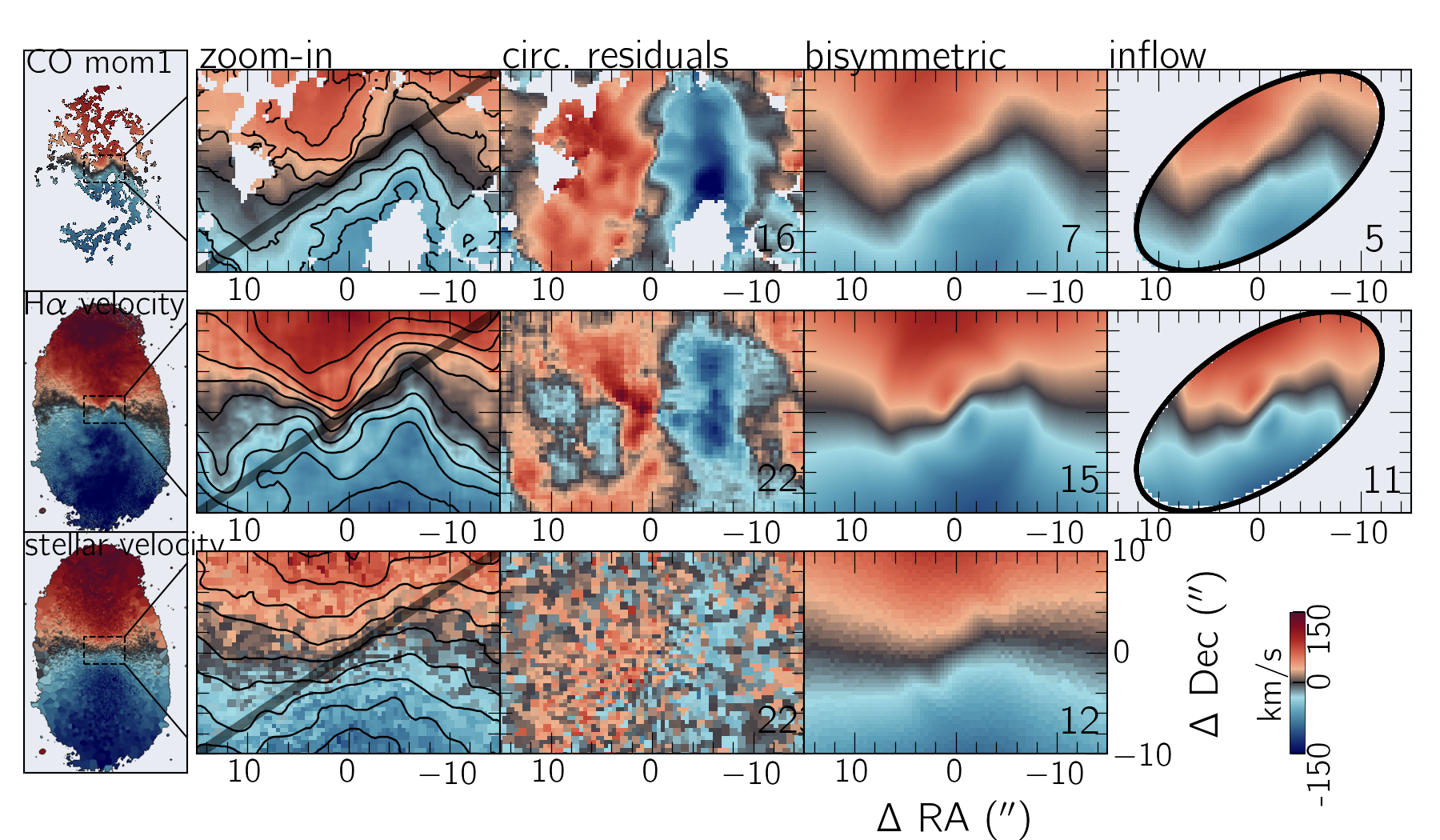}%
  \caption{{\it First column:} Line-of-sight velocities of CO, \ha~and the stellar velocity in NGC~1087. {\it Second column:} Zoom-in around the bar-structure, with $\pm25$~\kms~spaced iso-velocity shown on-top; the black straight line shows the P.A. of the faint bar at $\phi_{bar}^{\prime}=305^{\circ}$. {\it Third column:} residual velocities after subtracting the circular rotation in each velocity map. {\it Fourth column:} Bisymmetric model after fixing the oval orientation at $\phi_{bar}^{\prime}$. {\it Fifth column:} inflow model assuming the gas streams along $\phi_{bar}^{\prime}$; the individual radial profiles and expressions are shown in Figure~\ref{fig:all_vels}. The root mean square of the models is shown at the bottom right of each panel. All maps share the same color-bar except for the residual maps which cover $\pm50$~\kms~following the same colors scheme.}
  \label{fig:vel_maps} 
\end{figure*}

\subsection{Bar light modeling}
\label{sec:light_profiles}
The method adopted here is similar to the non-parametric method from \citet{Reese2007}, which is the same algorithm used by \XS~\citep[e.g.,][]{XookSuut} for extracting the different velocities in the kinematic models, as we explain in the following sections. Our non-parametric model minimizes the function $\chi^2 = (I_{obs}-I_k(r) w_k(r))^2$, where $I_{obs}$ is the observed intensity map, $I_k$ represents a set of intensities that will be estimated at different annuli, and $w_k$ is a set of weighting factors that will serve for performing a linear interpolation between the estimated $I_k$ intensities. The implementation on the F200W and F300M band filters is shown in Figure~\ref{fig:xs_phot}.

The NIR light from the bar-structure can be successfully modeled by an elliptical light distribution with with a constant orientation in the sky. The P.A. of such ellipse is
$\phi_{bar}^{\prime}=305^{\circ}$\footnote{Sky angles relative the disk major axis can be translated to angles measured in the galaxy plane and vise versa through, $\tan \phi_{bar} = \tan \Delta \phi / \cos i$, with $\Delta \phi = \phi_{bar}^{\prime} - \phi_{disk}^{\prime}$.} and $\varepsilon^{\prime}_{bar} = 0.5$, as shown in Figure~\ref{fig:xs_phot}. From this analysis the estimated 
semi-major axis length of the NIR bar is $a_{bar}^{\prime} = 14\arcsec$ or $18\arcsec \sim 1.2$\,kpc\footnote{Distances measured at different P.A. in the sky plane can be translated to distances in the galaxy plane following: $r_{disk}^2 = r_{sky}^2 (\cos^2 \Delta \phi + \sin^2 \Delta \phi / \cos^2 i)$.} on the disk at the considered distance. However, this structure is very diffuse, with most of the 2$\mu$m emission $\sim70\%$,  concentrated within the inner $6\arcsec\sim500$~pc. The latter region encloses the two bright spots observed in the MUSE optical continuum images. 

\section{Ionized and molecular gas}
The central panel of Figure~\ref{fig:main}  shows the ionized gas distribution traced by the \ha, \sii~and \oiii~emission-line fluxes. 
In general, galaxies hosting stellar bars do not frequently present cold or warm gas along the bar. This is often explained as a lost of angular momentum after the gas encounters the offset ridges, resulting in an infall of gas towards a central ring \citep[e.g.,][]{Athanassoula1992}.
NGC~1087 shows plenty of ionized gas throughout the disk and within the bar region. Furthermore, as we will see later, multiple enhanced CO bloops are spatially coincident with the F200W emission. 
To investigate the dominant ionizing source along the bar-like structure we made use of line-ratios sensitive to the ionization, in particular the \oi6300/\ha~and \oiii/\hb. The \oi6300/\ha~ratio has the advantage of being a good indicator of the presence of shocks.
The top panel from Figure~\ref{fig:bpt} shows these line-ratios color-coded with the $2\mu$m emission.
{ The observed ionized gas} is compatible with being produced by star-formation (SF) according to the \cite{Kewley2006} demarcation line.
However, dust lanes in bars are often associated with shocks as a result of the
complex gas dynamics \citep[][]{Athanassoula1992}. The bar-region in NGC\,1087 shows several dust lanes, thus we adopt shock models to investigate whether shock ionization can reproduce the observed line ratios. For this purpose we used the photoionization grids from MAPPINGS V~\citep[e.g.,][]{mappingsv} computed by \cite{Alarie2019}; we adopt the \cite{Gutkin2016} metallicities.
The latter parameter was set from the oxygen gas-phase abundance, adopting the \cite{Pilyugin2016} calibrator based on strong emission-lines (see Figure~\ref{fig:metal}). For the bar-region we find $\mathrm{ 12+\log O/H \sim 8.5}$ corresponding to $\mathrm{ Z_{ISM} = 0.008 }$ in the \cite{Gutkin2016} models.
For this metallicity, the predicted emission-line intensities from shock models can not explain the observed \oi/\ha~line ratios as observed in the top panel of Figure~\ref{fig:bpt}.
Thus, if shocks are happening in the bar-region, then their optical emission is not expected to be the dominant ionizing source. Although there is still a possibility that a combination of SF plus shocks could contribute to the observed line-ratios \citep[e.g.,][]{Davies2016}.

Additionally, Figure~\ref{fig:bpt} shows the equivalent width of \ha~(EW(\Ha)) for the ionized gas in the bar-region. Hot, old, and low-mass evolved stars are characterized for producing EW(\Ha)$~\lesssim 3$\AA~ \citep[e.g.,][]{Stasinska2008,Cid2010,Lacerda2018}.
The large values of EW(\Ha) $\gtrsim 100$\,\AA~found in the bar, can not be explained by the ionizing continuum emitted by this population of stars. 
Therefore, if most of the $2\mu$m emission observed along the bar is due to old stars, then their ionizing continuum is not sufficient to explain the observed line-ratios.  

The star formation rate (SFR) was estimated from the \ha~luminosity \citep[][]{Kennicutt1998}, after correcting the \ha~flux from dust extinction adopting the 
\cite{Cardelli1989} extinction law with $\mathrm{R_V = 3.1}$,  and assuming an intrinsic flux ratio of \ha/\hb = 2.86 and ionized gas temperature of T $\sim10^4$\,K corresponding to case B recombination \citep[e.g.,][]{Osterbrock}. The integrated SFR
within the NIR bar (i.e., Figure~\ref{fig:xs_phot}) is  $\mathrm{SFR(H\alpha)_{bar}} = 0.08~\mathrm{M_{\odot}/yr}$, while the total  $\mathrm{SFR(H\alpha)_{total}} = 0.4~\mathrm{M_{\odot}/yr}$.
The specific star formation rate (sSFR) was obtained from dividing the SFR surface density (=SFR/pixel area) by the stellar mass surface density as shown in Figure~\ref{fig:ssfr}. We note a clear enhancement in SF along the bar main axis; furthermore, the molecular surface density shows an enhancement in the same region.

Overall, our analysis suggests that the ionized gas in the bar structure is mostly associated to SF processes. Following we investigate whether the CO and SF concentration is induced by the bar potential.
\section{Inner Oval distortion}
If a perturbation in the gravitational potential, such as that induced by a bar potential, { is causing the} $2\mu$m light distribution to elongate along a preferred direction; then the particles in such structure are expected to follow quasi-elliptical orbits \citep[e.g.,][]{Athanassoula1992}. Hence kinematics of gas and stars should reflect this perturbation \citep[][]{Pence1984b}.

Figure~\ref{fig:vel_maps} shows the CO, ionized and stellar kinematics around the bar region. The gas, being collisional, is more sensitive to non-axisymmetric perturbations. This is clearly reflected in the CO~moment~1 map where a strong distortion of the semi-minor axis is observed.
The former misalignment is a signature of an oval distortion produced by a bar \citep[e.g.,][]{Pence1988,LopezCoba2022}, here observed at high spatial resolution.
The distortion is also observed in the \ha~velocity map, however, since the ionized gas traced by \ha~is hotter ($\mathrm{T\sim10^4}$\,K), and therefore it presents a larger intrinsic velocity dispersion than the one of the molecular gas traced with CO ($\mathrm{T\sim10}$\,K), it is less pronounced in the ionized gas and is probably affected by local SF. 
The stellar kinematics although affected by the pixel-coadding during the SSP analysis, it { still reveal} a clear distortion in the iso-velocities { near the bar} as noted in Figure~\ref{fig:vel_maps}.

{ So far, it is clear} that the stars, the molecular gas traced with CO and the ionized gas traced with \ha~respond in a similar way to the presence of this bar-like structure detected in NGC~1087.

\subsection{Kinematic interpretation of the oval distortion}

The distortion in the velocity field such as the ones observed before could be caused by an elongated potential. 
However, only a few kinematic models in the literature attempt to describe the flow caused by an oval distortion \citep[e.g.,][]{ velfit, Maciejewski2012}.
A bisymmetric distortion induced by a second order perturbation to the potential, such as that induced by an elongated potential, has been shown to successfully reproduce the velocity field in bars \citep[e.g.,][]{velfit}. 
Since the F200W image evidences the presence of a faint bar-like structure oriented at $305^{\circ}$ in the sky, we performed bisymmetric models over the stellar, \ha~and CO velocity maps, with a fixed orientation of the oval distortion.

The bisymmetric model from \citet[][]{velfit} is described by the following expression:

\begin{multline}
 \label{Eq:bisymmetric}
 V_\mathrm{LoS}  =  \sin i \Big( V_t(r)\cos \theta - V_{2,t}(r)\cos 2\theta_\mathrm{bar} \cos \theta \\ - V_{2,r}(r) \sin 2\theta_\mathrm{bar} \sin \theta \Big) + V_\mathrm{sys} 
\end{multline}
where $\theta_{bar} = \theta - \phi_{bar}$; with $\theta$ being the azimuthal angle measured on the disk plane from the line of nodes and $\phi_{bar}$ the position angle of the bar-like structure on the disk plane. $V_{t}$ is the tangential or circular rotation and $V_{2r}$ and $V_{2t}$ represent the radial and tangential velocities that result from a bisymmetric distortion to the gravitational potential.

Before this analysis we performed a circular rotation model to obtain the disk projection angles and the rotational curves.
In this case and in the subsequent, we used the \XS~code for generating the kinematic models \citep[e.g.,][]{XookSuut}.
%.
This code derives an interpolated model over a set of concentric rings evenly spaced $r_k$, minimizing the function $\chi^2 = (V_{obs}-V_k(r) w_k(r))^2/\sigma^2$, where $V_{obs}$ is the observed velocity map; $V_{k}$ are the set of velocities inferred at $r_k$; $w_k$ is a set of weights that depend on the specific kinematic model adopted and will serve to create a 2D model; $\sigma$ is the error velocity map. We refer to \citet[][]{XookSuut} for a thorough description of this analysis. It is important to mention that the modeling does not assume any parametric function on the velocity profiles.

A pure circular rotation, without non-circular motions, is described by the first term on the right from Equation~\ref{Eq:bisymmetric}. 
The disk orientation of NGC\,1087 was estimated from modeling the \ha~and stellar velocity maps with circular rotation only, obtaining $\phi_{disk}^{\prime} = 358.9^{\circ}$, $i=44.5^{\circ}$ and $V_{sys}=1526$~\kms. 
The residual velocity maps from this model i.e., $V_{obs}-V_{circular}$, are shown in the third column from Figure~\ref{fig:vel_maps}.
The circular rotation models leave large-scale residual velocities of the order of $30$~\kms~with a mirror symmetry about the nucleus; this corresponds to  de-projected  amplitudes of $\sim 50$\,\kms~for the non-circular velocities on the disk plane. These residual patterns have been observed in larger scales in galaxies with long bars in \ha~\citep[e.g.,][]{Lang2020, LopezCoba2022} and molecular gas \citep[e.g.,][]{Pence1988, Mazzalay2014}.

The bisymmetric model with a fixed oval orientation at $\phi_{bar}^{\prime}$ is shown in the fourth column in Figure~\ref{fig:vel_maps}. This model reproduces simultaneously the twisted iso-velocities in the three velocity maps. The root mean square (rms) of the models decrease compared with the circular rotation one, therefore, in terms of the residuals, the LoS-velocities observed around the bar region can be reproduced by a kinematic model that considers a bar-like perturbation as the main source of non-circular motions. However, at such high spatial resolution $\sim 70$\,pc, there are still residual velocities that the bisymmetric model can not account for, this is reflected in the 15~\kms~rms in the models, which is similar or larger than the level of turbulence of the ISM \citep[e.g.,][]{Moiseev2015}. 

An alternative interpretation to the observed non-circular motions in the \ha~and CO velocity maps is the presence of gas inflow induced by the NIR-bar \citep[][]{Mundell1999}; in fact, hydro-dynamical models simulating bars predict inflow of gas along the offset ridges, or bar dust lanes, \citep[e.g.,][]{Athanassoula1992}. As observed in Figure~\ref{fig:main}, the MUSE image of NGC\,1087 shows several dust lanes in the central region making difficult the identification of those associated with the bar.
Assuming the spiral arms are trailing (see bottom-right panel from Figure~\ref{fig:main}), NGC~1087 rotates counterclockwise in the sky. Thus, positive (negative) residuals in the near (far) side represent inflow; hence, a radial flow ($V_{rad}$) along the bar major axis is inflowing if $V_{rad}<0$.
We implemented a non-axisymmetric model, with a flow streaming along $\phi_{bar}^{\prime}$.
The non-axisymmetric inflow model is described by the following expression:
\begin{multline}
 \label{Eq:inflow}
 V_{LoS} = (V_{t}(r)\cos\theta + V_{rad}(r)\cos \theta_{bar} \sin \phi_{bar}) \sin i \\ + V_{sys} 
\end{multline}
where $V_{rad}$ is the radial velocity of the flow, and $\phi_{bar}$ is the misalignment between the disk and the bar position angles on the disk plane. This model assumes the gas is flowing along $\phi_{bar}$. This expression is similar to \cite{Hirota2009} and \cite{Wu2021} assuming that the gas flows parallel to the bar major axis. 

\begin{figure}[t!]
  \centering
    \includegraphics[]{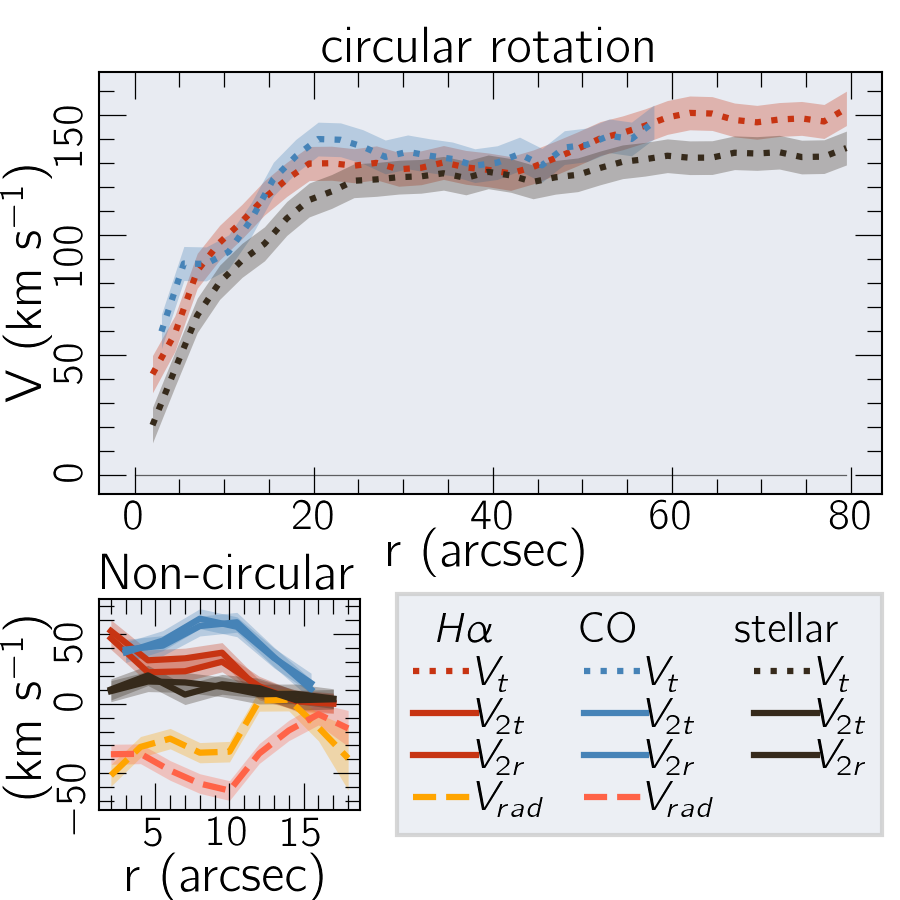}%
  \caption{Radial distribution of the different velocity components for the considered kinematic models. $V_t$ is the disk pure circular rotation, or rotational curve; $V_{2r}$ and $V_{2t}$ are the bisymmetric velocities; $V_{rad}$ the radial flow from the non-axisymmetric flow model. In all non-circular models the position angle of the oval distortion was fixed to $\phi_{bar}^{\prime} = 305^{\circ}$ (namely, $297^{\circ}$ in the galaxy plane). Lines in red colors represent results of models computed on the \ha~velocity map, blue lines on the CO moment~1 and black lines on the stellar velocity map. Shaded regions represent $1\sigma$ errors. The non-circular velocities were estimated up to $18\arcsec$, which covers the de-projected length of the bar. A $2\arcsec$ sampling step was adopted, i.e., larger than the FWHM spatial resolution of the data.}
  \label{fig:all_vels}
\end{figure}

In order to consider pixels likely affected by this motion, we consider the elliptical region that better describes the bar-like light distribution and defined in Figure~\ref{fig:xs_phot}.
The 2D representation of this model is shown in Figure~\ref{fig:vel_maps}, while the 
kinematic radial profiles of all models are shown in Figure~\ref{fig:all_vels}. As observed, the radial velocity $V_{rad}$ results negative in the model, with maximum inflow velocities of the order of 50\,\kms.

\begin{figure}[t]
\centering
    \includegraphics[]{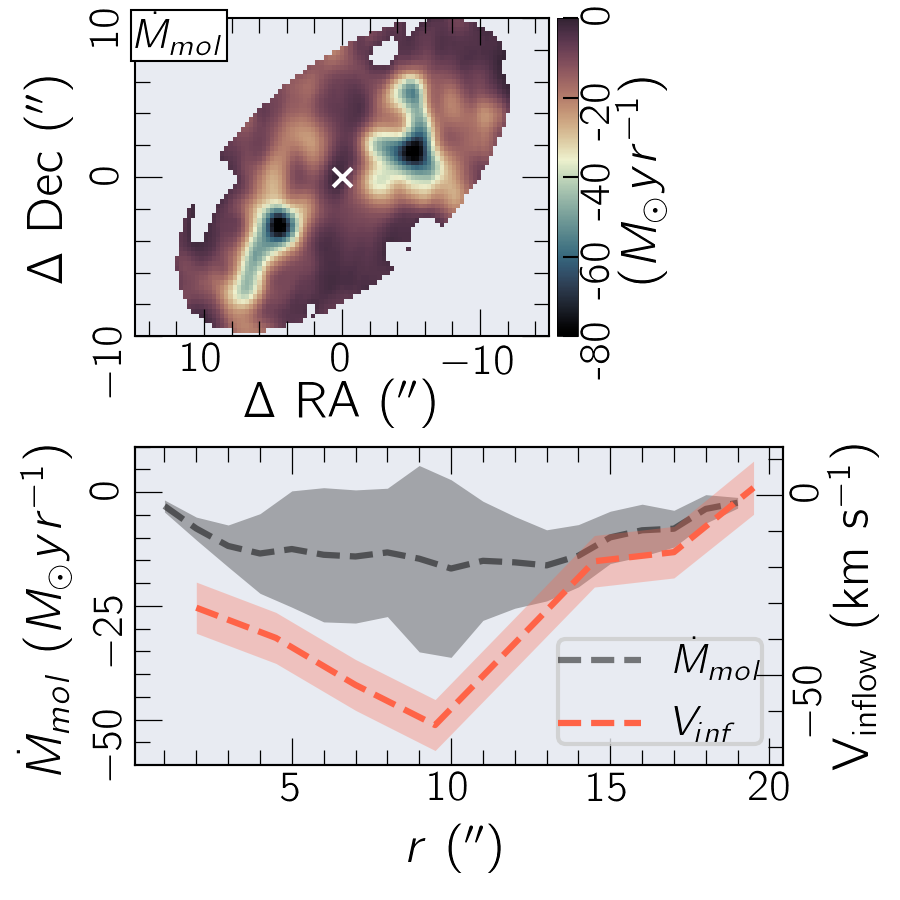}%
  \caption{Molecular mass inflow induced by the bar-like structure. {\it Top panel:} 2D distribution of $\dot{M}_{mol}$. {\it Bottom panel:} Average radial distribution of $\dot{M}_{mol}$ in black and radial velocity profile of the inflow velocity in red colors. Shadow black regions represent the standard deviation of $\dot{M}_{mol}$ at each radial bin. A $\mathrm{SN>3}$ was applied to the moment~0 map to exclude spurious CO detection.}
  \label{fig:inflow}
\end{figure}

The molecular mass flow rate ($\dot{M}_{mol}$) associated can be computed following the expression \citep[e..g,][]{DiTeodoro2021}: 
\begin{equation}
 \dot{M}_{mol}(r) = 2\pi r \Sigma_{mol}(r) V_{inflow}(r)
\end{equation}
with $\mathrm{ \Sigma_{mol} }$ being the de-projected molecular mass surface density, $V_{inflow}$ the CO inflow velocity and $r$ the galactocentric distance. Figure~\ref{fig:inflow} shows the spatially resolved $\dot{M}_{mol}$ and its average radial profile. We find an average $\dot{M}_{mol}\sim -20 \mathrm{M_{\odot}/yr}$. For comparison, spiral arms induce radial flows and radial velocities of the order of $1\mathrm{M_{\odot}/yr}$ and $10$~\kms~respectively, \citep[e.g.,][]{DiTeodoro2021}.

\section{Discussion and Conclusions}
The combination of spatially resolved spectra provided by MUSE-VLT, with the high spatial resolution from JWST and the ALMA CO~$J=2\rightarrow 1$ data allowed us to reveal a central kinematic oval distortion, as well as a small scale elongated structure in the NIR continuum in NGC~1087. 
This elongated structure is not evident in optical images, but only revealed with the NIRCam F200W and F360M filters at $2\mu$m and $3\mu$m respectively, thanks to its exquisite resolution. The optical counterpart of this bar is likely to be affected by dust absorption and by the intense radiation from young stars.
Its preferential elongation suggests the presence of a non-axisymmetric perturbation in the gravitational potential, in particular a bar-like type. At first order, the 2$\mu$m  light distribution of this structure can be described with an exponential disk profile oriented at $\phi_{bar}^{\prime} = 305^{\circ}$, and the ellipticity of the elongated structure is $\varepsilon^{\prime} = 0.5$, (see Figure~\ref{fig:xs_phot}), with a half-major axis length of $a_{bar}^{\prime}\sim 14\arcsec$ or $\sim$~1.2~kpc in the galaxy plane.
Overall the shape of this structure resembles to a faint stellar bar, although with remarkable differences with respect to conventional stellar bars, in terms of bar-length, the lack of a SF nuclear ring and not clear associated dust lanes along the bar axis. The same argument however applies for an elongated bulge. The main reason this is not considered is due to the large reservoirs of molecular gas observed, since bulges tend to be quiescent structures with null or low SFRs \citep[][]{Shapiro2010}.   
Furthermore, the \oi$/$\ha~and \oiii$/$\hb~line-ratios and large EW\ha~from the bar (see Figure~\ref{fig:main}) indicate that the ionized gas is product from a recent or ongoing SF event given its spatial coincidence with the CO emission.

The ionized and molecular gas kinematics and the stellar velocity, reveal simultaneously the presence of an oval distortion in the velocity maps in concordance with the extension of the faint bar.
This evidences undoubtedly the imprints of a non-axisymmetric perturbation in the potential, here captured in different phases and enhanced in dynamically cold gas.

%note how the $0$~\kms~iso-velocity twist is more pronounced in CO rather than in \ha~or in the stellar velocity.

Kinematic models for a second order potential perturbation successfully described the  LoS velocities in the three velocity maps. We showed that this is achieved just by fixing the orientation of the stream flow to the bar major axis in the models, and the radial profiles of the noncircular velocities are consistent with the residuals from circular rotation, with possible differences in amplitudes due to assymmetric drift and other random motions not considered in the model as we showed in Figure~\ref{fig:all_vels}. 
{ Apart} from the elliptical motions considered in the bisymmetric models, radial inflows are expected in bars, and these are observed to occur along the bar-dust lanes \citep[][]{Wu2021,Sormani2023}.
Our implementation of an inflow model is therefore physically motivated.
An inflow of cold gas caused by a lose of angular momentum  could trigger the SF observed along the bar. Although a fraction of ionized gas could arise by the shock with the dust lanes, the contribution of shocks is expected to be minor.
Moreover, the sign of the radial flow suggests the gas is inflowing to the center at a maximum speed of $50$~\kms. This kinematic model { yields} the lowest rms { despite being} the model with the lowest number of free variables. Statistically speaking, the inflow scenario is favored over the bisymmetric model, although given the complex behavior of gas in bars, the contribution of gas oval orbits can not be ruled out. In fact both type of motions are expected to happen simultaneously in bars \citep[][]{Athanassoula1992, Regan1999}. To date, there is no a 2D-kinematic model that includes such dynamics of bars.
The  gas inflow model is in concordance with the observed enhancement in \ha~flux along the bar. The inferred inflow velocities translate in molecular mass inflow rates of $\dot{M}_{mol} = 20~ \mathrm{M_{\odot}/yr}$. This is larger than the measured SFR(\ha) along the bar, thus we argue that a major inflow of gas might be feeding the SF in the bar. In fact SF prevents the so-called  continuity problem \citep[e.g.,][]{Simon2003,Maciejewski2012} when often considering radial flow models. This large inflow rate is partially due to the large inflow velocities; although this amplitude of non-circular motions is observed in bars, the global inflow rate expected by hydrodynamical models is lower \citep[][]{Athanassoula1992,Mundell1999}.

To summarize, in this work we have taken the advantage of multiple archival data to reveal a hidden non-axisymmetric structure in the optical, thanks to the high angular resolution of the data. We were able to confirm the presence of an oval distortion and successfully model it with a bar-like flow model. We also show that the SFR in the bar could be explained by an inflow of gas along the bar major-axis. Our results contribute to understand the overall picture of non-axisymmetry in galaxies with the most sophisticated data so far. Additionally, it highlights the importance of using the IR instead of optical bands to detect stellar bars, which could increase the fraction of barred galaxies detected in the local Universe.

\section*{Acknowledgment}
We thank the anonymous referee for their comments and suggestions which help to improve the quality of this manuscript.
CLC acknowledges support from Academia Sinica Institute of Astronomy and Astrophysics. LL acknowledges the Ministry of Science \& Technology of Taiwan under the grant NSTC 112-2112-M-001-062 -.

Based on data obtained from the ESO Science Archive Facility with DOI(s): \dataset[https://doi.org/10.18727/archive/47]{https://doi.org/10.18727/archive/47}.
The specific PHANGS–JWST observations analyzed can be can be found in MAST: 
\dataset[https://doi.org/10.17909/9bdf-jn24]{https://doi.org/10.17909/9bdf-jn24}.
This paper makes use of the following ALMA data: ADS/JAO.ALMA\#2018.1.01651.S.

%

%%%%%%%%%%%%%%%%%%%%%%%%%%%%%%%%%%%%%%%%%%%%%%%%%%



\bibliographystyle{aasjournal}

\end{document}